\documentclass[aip,
nofootinbib,
rsi,%
reprint,
twocolumn,
longbibliography,
author-numerical%
]{revtex4-2}

\usepackage{amsfonts,amssymb,amsmath}
\usepackage[usenames]{color}
\usepackage{graphicx}
\usepackage{bm}
\usepackage{changes,comment}

\newcommand{\pdr}[2]{\dfrac{\partial{#1}}{\partial {#2}}}
\newcommand{\pddr}[2]{\dfrac{\partial^2{#1}}{\partial {#2}^2}}

\newcommand{\pdra}[2]{{\partial{#1}}/{\partial{#2}}}
\newcommand{\pddra}[2]{{\partial^2{#1}}/{\partial{#2}^2}}

\newcommand{\tx}{\tilde{x}}

\newcommand{\tk}{\tilde{k}}

\newcommand{\tj}{\tilde{j}}

\newcommand{\teta}{\tilde{\eta}}
\newcommand{\bars}{\bar{s}}

\newcommand{\tD}{\widetilde{D}}

\newcommand{\tp}{\tilde{p}}

\newcommand{\tl}{\tilde{l}}

\newcommand{\ksig}{k_{\sigma}}
\newcommand{\tpox}{\tp_{O_2}}
\newcommand{\poxref}{p_{O_2}^{ref}}

\newcommand{\cref}{c_{ref}}
\newcommand{\Cdl}{C_{dl}}
\newcommand{\Alam}{A_{\lam}}

\newcommand{\lam}{\lambda}

\newcommand{\Dox}{D_{ox}}

\newcommand{\lexp}[1]{\exp\left(#1\right)}

\newcommand{\sion}{\sigma_p}

\newcommand{\lcat}{l_t}

\newcommand{\tom}{\tilde{\omega}}
\newcommand{\ri}{{\rm i}}
\newcommand{\expo}{{\rm e}^{\teta^0}}

\newcommand{\etal}{{ }et al.{ }}

\begin{document}


\sf

\title{A model for water transport in the membrane and an impedance spectroscopy study
of the effect of relative humidity on PEM fuel cell parameters}

\author{Andrei Kulikovsky}
\thanks{ECS member}
\email{A.Kulikovsky@fz-juelich.de}

\address{Forschungszentrum J\"ulich GmbH           \\
    Theory and Computation of Energy Materials (IET--3)   \\
    Institute of Energy and Climate Research,              \\
    D--52425 J\"ulich, Germany
}

\author{Tatyana Reshetenko}
\thanks{ECS member}
\email{tatyanar@hawaii.edu}

\address{Hawaii Natural Energy Institute \\
    University of Hawaii \\
    Honolulu, Hawaii 96822, USA
}

\date{\today}

\begin{abstract}
Effective water management is essential for the optimal performance of PEM fuel cells.
We have developed an impedance model for liquid water transport through the membrane
and coupled it with the two-phase model for cathode side impedance.
The complete model was fitted to experimental spectra measured
at anode/cathode relative humidities (RH) of 32/32\%, 50/50\% and 100/100\%
within a current density range of 100 to 1000 mA cm$^{-2}$ and an air flow stoichiometry
of 2. Cathode catalyst layer (CCL) saturation decreases with current density
due to a growing liquid pressure gradient.
For all RH values, the CCL oxygen diffusivity increases dramatically with cell current
due to progressive involvement of larger pores into the proton current conversion.
Higher RH leads to higher double layer capacitance, which indicates that liquid water
increases the electrochemically active surface area.
\end{abstract}


\maketitle


\section{Introduction}

Water management is one of the critical issues in optimizing PEM fuel cells performance.
Polymer electrolyte in the cathode catalyst layer (CCL) conducts protons only in a wet state.
On the other hand, excess of liquid water may severely retard
oxygen transport to reaction sites. Water in the CCL arrives from the
anode through the membrane and produced in
the oxygen reduction reaction (ORR). The balance of water fluxes in the CCL
thus depends on the current density, which greatly complicates the problem.

Water management in PEMFCs porous layers has been the subject of numerous experimental
and modeling studies (see the review~\cite{Liu_22}).
On the modeling side, the problem has been studied using two-phase
lattice-Boltzmann modeling (LBM)~\cite{Mukherjee_09,Nazemian_18,Shimpalee_20,Grunewald_21,Chen_21},
pore-network modeling (PNM)~\cite{Hannach_12,Wu_12,Gorp_23},
and macrohomogeneous modeling~\cite{Eikerling_06a,Olbrich_22b,Kulikovsky_24b}.
The most accurate data could potentially be obtained from LBM and PNM,
which employ synthetic or reconstructed from experiments porous structures.
However, accurately reproducing the CCL structure, which includes carbon particles,
platinum species, and ionomer, remains challenging. Furthermore, LBM and PNM are
very time-consuming and are therefore not suitable for fitting impedance spectra.

The slowest processes in PEM fuel cells are the dynamics of liquid water in
the polymer electrolyte membrane. The water flux through the membrane
is the sum of the electroosmotic flux directed from the anode to the cathode,
and the back diffusion flux, which transports water in the opposite direction.
The net water flux through the membrane alters
the CCL liquid saturation, impacting the CCL impedance.

Electrochemical impedance spectroscopy (EIS) is a powerful tool for analyzing
transport processes in PEMFC porous layers~\cite{Lasia_book_14}.
To extract cell transport and kinetic parameters, a relevant
physics-based model must be fitted to the experimental spectra.
In principle, any transient model of cell performance can be
linearized and Fourier-transformed for impedance calculations.
However, the key issue for spectra fitting is the speed of model calculations.

One of the first steady-state PEMFC performance models that took into account
water transport in the membrane was developed by
Springer, Zawodzinski and Gottesfeld~\cite{Springer_91}.
The model employed water transport parameters (drag and water diffusion
coefficients) that had been measured by the authors. Five years later,
Springer etal\cite{Springer_96}  developed an impedance model of the cell.
However, this model ignored water transport in the membrane.

Models of two-phase water transport in the cell
vary from one-dimensional through-plane models~\cite{Chen_07,Olapade_11}
to multidimensional CFD-based models, which take into account the complex geometry of cells~\cite{Natarajan_01,Wang_07a,Gurau_08,Berg_10,Berning_11}.
However, very few papers on PEMFC impedance modeling have so far
included water dynamics in the membrane.
Bao and Bessler calculated the cell impedance using their
two-dimensional, two-phase, transient CFD model of PEM fuel cell performance~\cite{Bao_15}.
They calculated the response of the cell current to a step-like change
of the cell potential and performed a Fourier transform
of the applied and response signals to obtain the impedance.
To the best of our knowledge, Ref.\cite{Bao_15} is
the only work in which the cell impedance was calculated based
on a physics-based performance model. However, this time-consuming
calculation is not suitable for spectra fitting.

In this work, we have developed a one-dimensional (1d)
model for the impedance of the membrane separating
the PEMFC anode and cathode. The model is based on a transient
equation for water transport through the membrane. The model
is coupled with a two-phase impedance model of the cell cathode
side~\cite{Kulikovsky_24b,Kulikovsky_24e}.
We fitted the complete impedance model to eighteen PEM fuel cell spectra,
which were measured at anode/cathode relative humidities
of 32/32\%, 50/50\% and 100/100\%, and
current densities $j_0$ of 100, 200, 400, 600, 800 and 1000 mA~cm$^{-2}$.

Contrary to the common belief, with the growth of $j_0$, the CCL mean liquid
saturation $\bars$ decreases. This is due to the gradient of the liquid water
pressure in the CCL, which increases with the current and effectively removes
liquid water from the catalyst layer. The CCL oxygen diffusivity $\Dox$
increases dramatically with the cell current. This growth is partly due to
the reduction in the CCL liquid saturation. However, the main mechanism behind
this increase is the progressive involvement of larger pores in
proton current conversion\cite{Kulikovsky_24e}.

The highest RH of 100/100\% provides the highest double layer capacitance
and the lowest bulk membrane resistivity. The ORR Tafel slope increases with the
cell current, showing significantly larger values observed at
32/32\% RH as compared with the other two RH combinations.


\section{Model for membrane impedance}

\subsection{Model assumptions and basic equations}

The model is based on the following assumptions:

\begin{itemize}

\item The cell is isothermal, i.e., the water flux in
the membrane due to a temperature gradient is neglected.

\item The electroosmotic water drag coefficient $n_d$ in the membrane is constant.

\item The gaseous pressure in the CCL is constant. Water
evaporation/condensation in the CCL is neglected.

\item The membrane water content on the cathode side is the sum of the equilibrium
   and dynamic terms, as discussed below.

\item The static water flux entering the cathode catalyst layer from the
   membrane is $\alpha_w j_0$, where $\alpha_w = 0.2$ is the net water transfer
   coefficient and $j_0$ is the cell current density.

\end{itemize}
These assumptions are discussed in the text when they are first used.

The liquid water flux $N_w$ in the membrane comprises the diffusion and electroosmotic
components:
\begin{equation}
   N_w = - D_w \pdr{c_w}{x} + \dfrac{n_d j_0}{F}
   \label{eq:Nm}
\end{equation}
where
$c_w$ is the liquid water molar concentration,
$j_0$ is the cell current density,
$D_w$ is the membrane liquid water diffusivity, and
$n_d$ is the drag coefficient, a number of water molecules transported by one proton.

In the membrane, no charges are produced and hence $\pdra{j_0}{x} = 0$.
The continuity equation for $c_w$ is thus
\begin{equation}
   \pdr{c_w}{t} - \pdr{}{x}\left(D_w \pdr{c_w}{x}\right) = 0
   \label{eq:cmx}
\end{equation}

\begin{figure}
\begin{center}
   \includegraphics[scale=0.8]{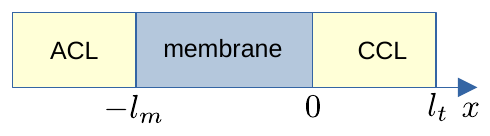}
\end{center}
\caption{Schematic of the cell layers and the $x$--coordinate.
}
\label{fig:sketch}
\end{figure}

To simplify the coupling of the membrane model with the
CCL performance model, we fix the origin of the $x$--axis
at the membrane / CCL interface. The membrane problem is, therefore,
formulated in the domain $x \in [-l_m, 0]$, where $l_m$ is the
membrane thickness (Figure~\ref{fig:sketch}).

It is convenient to reformulate Eq.\eqref{eq:cmx} in terms of a number of
water molecules  $\lam$ per sulfonic group in the membrane:
\begin{equation}
    \lam = \dfrac{W_m c_w }{\rho_m}
    \label{eq:lam_def}
\end{equation}
where $W_m$ is the ionomer equivalent weight and $\rho_m$ is the mass density of a dry ionomer.
With Eq.\eqref{eq:lam_def}, equation \eqref{eq:cmx} transforms to
\begin{multline}
   \pdr{\lam}{t} - \pdr{}{x}\left(D_w(\lam) \pdr{\lam}{x}\right) = 0, \\
     - \dfrac{\rho_m}{W_m} \left.D_w \pdr{\lam}{x}\right|_{x=- l_m}  + \dfrac{n_d j_0}{F}
              = - \left.\dfrac{K_L}{V_w \mu_w}\pdr{p_L}{x}\right|_{x=0+}, \\
        \lam|_{x=0-} = \Lambda(RH^c) + \alpha_s s|_{x=0+}.
   \label{eq:lamx}
\end{multline}
where the first boundary condition expresses equality of water fluxes
on either side of the membrane.
Here, $K_L$ is the hydraulic permeability of the CCL, $\mu_w$ is the liquid water
viscosity, and $V_w$ is the molar volume of water.

\begin{figure}
\begin{center}
   \includegraphics[scale=0.45]{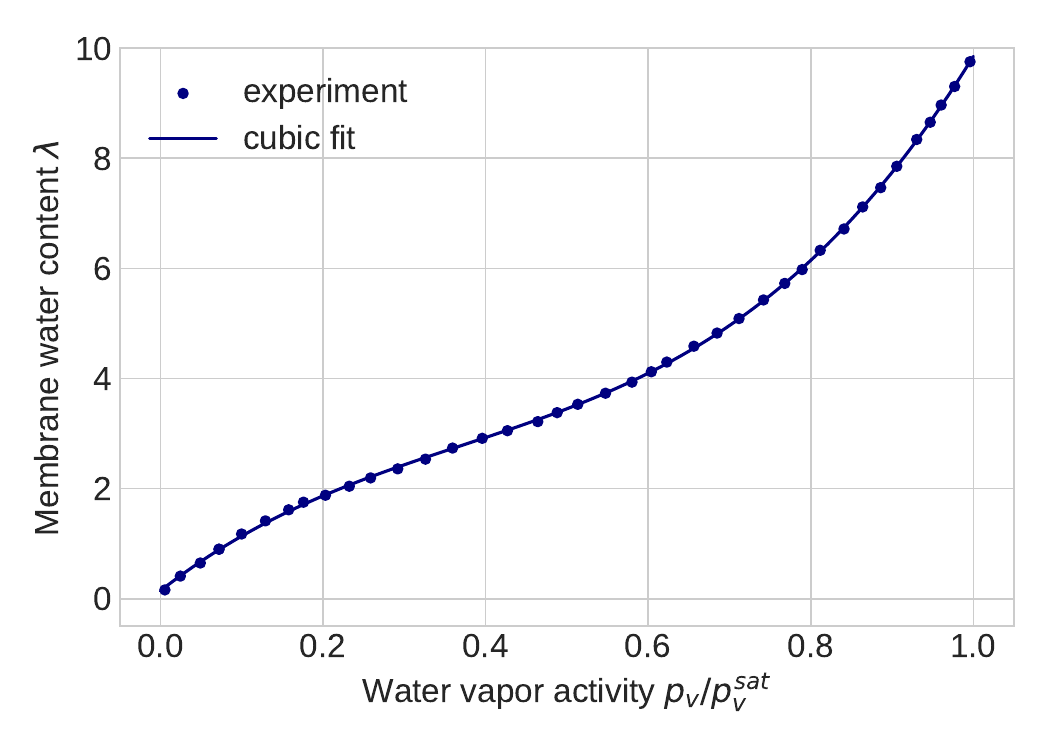}
\end{center}
\caption{Points -- the membrane water sorption isotherm measured by
   Hinatsu, Mizuhata and Takenaka \cite{Hinatsu_94},  solid line -- Eq.\eqref{eq:Lam}.
   The points were digitized from the best-fit curve provided in Ref.\cite{Hinatsu_94}
}
\label{fig:wup}
\end{figure}

The second (right) boundary condition for Eq.\eqref{eq:lamx} is a model
assumption meaning that the membrane
water content on the cathode side is the sum of two terms. The first term is
the static equilibrium water content which depends on the cathode RH via
the membrane water sorption isotherm $\Lambda(RH)$ (Figure~\ref{fig:wup}).
The second term is proportional to the CCL liquid water saturation. Here, $\alpha_s$
is a dimensionless coefficient on the order of 10, declared below
as a fitting parameter.
The reason behind this assumption is that the liquid water produced in the ORR
wets the CCL membrane phase, which is
in contact with the bulk membrane. The ionomer film in the CCL is thin, and its wetting
occurs much faster, than the characteristic time for water transport through the bulk
membrane. In other words, we assume that any variation in the CCL liquid saturation
at the membrane interface affects the water content of
the membrane cathode side immediately.

\begin{figure}
\begin{center}
   \includegraphics[scale=0.45]{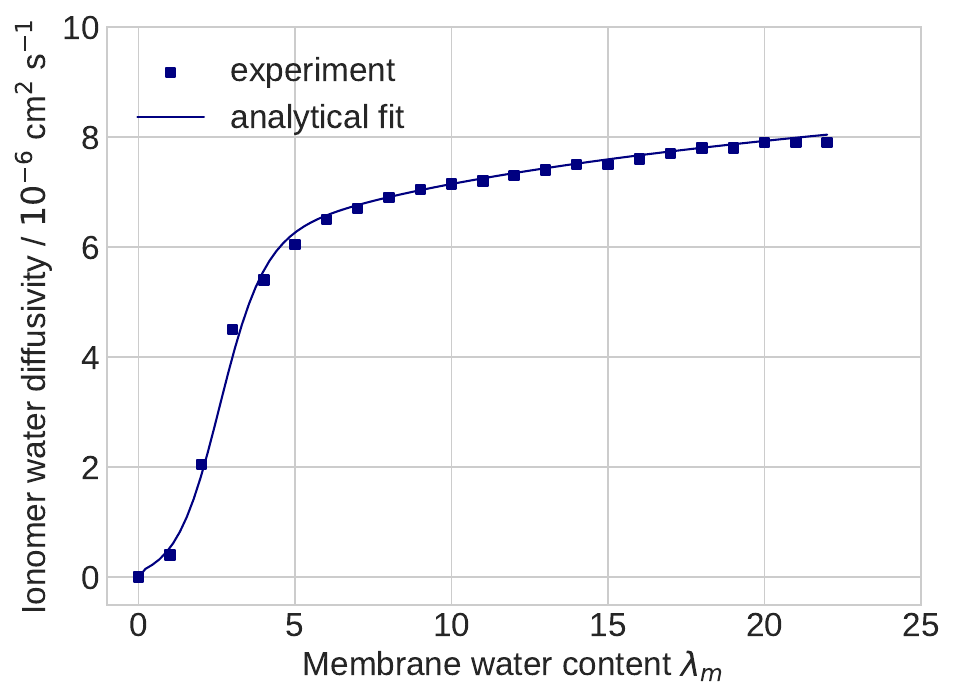}
\end{center}
\caption{Points -- the diffusion coefficient  $D_w$ of water
    in membrane measured by van Bussel \etal\cite{Bussel_98}. Solid line --
    the analytical fit, Eq.\eqref{eq:Dmfit}.
}
\label{fig:Dm}
\end{figure}

The feature of Eq.\eqref{eq:lamx} is the dependence of the diffusivity $D_w$
on $\lam_w$, which makes Eq.\eqref{eq:lamx} strongly nonlinear.
The plot of $D_w$ measured in experiments of van Bussel \etal\cite{Bussel_98}
is shown in Figure~\ref{fig:Dm} together with the analytical fit suggested in\cite{Kulikovsky_03a}:
\begin{multline}
    D_w  = 4.1\cdot 10^{-6} \left(\dfrac{\lambda}{25}\right)^{0.15}
          \left[ 1 + \tanh\left(\dfrac{\lambda-2.5}{1.4}\right)\right], \\
         \text{cm$^2$s$^{-1}$}
       \label{eq:Dmfit}
\end{multline}
Satterfield and Benziger measured dynamics of the membrane water adsorption
by placing a dry membrane in air fully saturated with water vapor~\cite{Satterfield_08}.
The characteristic time of membrane wetting was on the order of 1000 s~\cite{Satterfield_08}.
This result agrees qualitatively with
the $D_w$ curve in Figure~\ref{fig:Dm}. Indeed,  when the membrane is dry,
the characteristic time for water diffusion through the membrane $\tau_D \simeq l_m^2 / D_w$
tends to infinity.

The cubic fit for the Nafion membrane water uptake isotherm at 80$^\circ$C is
\begin{multline}
    \Lambda_{RH}(a) = 0.1446 + 11.57 a - 17.99 a^2 + 16.12 a^3
   \label{eq:Lam}
\end{multline}
where $a = p_v / p_v^{sat}$ is the water vapor activity  (Figure~\ref{fig:wup}).
Eq.\eqref{eq:Lam} fits the experimental data
of Hinatsu, Mizuhata and Takenaka \cite{Hinatsu_94}.
Eq.\eqref{eq:Lam} provides a better fit in the range of $0 \leq a \leq 1$, than
the fit \cite{Kulikovsky_03a}  suggested for the range $0 \leq a \leq 3$.


\subsection{Equation for the perturbation amplitude of the membrane water content}

The linearized and Fourier-transformed equation for the dimensionless
perturbation amplitude of the membrane water content is derived
in Appendix~A:
\begin{multline}
   \tD^0 \pddr{\lam^1}{\tx}
     + \left(2 \pdr{\tD^0}{\lam} \pdr{\lam^0}{\tx} \right)\pdr{\lam^1}{\tx} \\
          +  \left(\left(\pdr{\lam^0}{\tx}\right)^2 \pddr{\tD^0}{\lam}
             + \pdr{\tD^0}{\lam}\pddr{\lam^0}{\tx} - \ri\tom \chi^2 \right) \lam^1 = 0,
   \label{eq:dFlam1x2}
\end{multline}
Here,
\begin{multline}
   \tD = \dfrac{4 F \cref D}{\sigma_* b}, \quad \tx = \dfrac{x}{\lcat},
      \quad \tom = \omega t_*,  \quad \tj = \dfrac{j}{j_*}, \\
       \tl_m = \dfrac{l_m}{\lcat}
      \label{eq:dless1}
\end{multline}
where $t_*$ is the characteristic time of the DL charging, $j_*$
is the characteristic proton current density
\begin{equation}
   t_* = \dfrac{\Cdl b}{i_*}, \quad j_* = \dfrac{\sigma_* b}{\lcat},
   \label{eq:tast}
\end{equation}
and the dimensionless parameter $\chi$ is
\begin{equation}
   \chi = \sqrt{\dfrac{4 F \cref i_* \lcat^2}{\Cdl \sigma_* b^2}}.
   \label{eq:chi}
\end{equation}

The boundary condition to Eq.\eqref{eq:dFlam1x2} on the cathode side
of the membrane is
\begin{equation}
     \lam^1|_{\tx=0-} = \alpha_s s^1|_{\tx=0+}
   \label{eq:dbcFlam1}
\end{equation}
Eq.\eqref{eq:dbcFlam1} means that the perturbation $\lam^1$
changes in-phase with the perturbation
of the CCL liquid saturation $s^1$, as discussed above.
Below, we will see that the solution of Eq.\eqref{eq:dFlam1x2} is not needed
and the CCL problem is closed using the boundary condition \eqref{eq:dbcFlam1}.



\subsection{Steady-state problem for water transport in the membrane}

The steady-state problem for Eq.\eqref{eq:lamx} in the dimensionless form is
\begin{multline}
   \pdr{}{\tx}\left(\tD^0 \pdr{\lam^0}{\tx} \right) = 0, \quad
             \lam^0|_{\tx=0-} = \Lambda(RH^c) + \alpha_s s^0|_{\tx=0+} \\
          \left(- \Alam \tD^0 \pdr{\lam^0}{\tx}\right)_{\tx=-\tl_m}  + n_d\tj^0
          = - \left.B_L\pdr{\tp_L^0}{\tx}\right|_{\tx=0+}.
   \label{eq:lam0x}
\end{multline}
Here, the dimensionless parameters $\Alam$ and $B_L$ are
\begin{equation}
   \Alam = \dfrac{\rho_m}{4 W_m \cref}, \quad
   B_L = \dfrac{F K_L \poxref}{\mu_w V_w \sigma_* b}
   \label{eq:BLA}
\end{equation}
and the notations \cite{Kulikovsky_24b} are used.

Experiments of Yan, Toghiani and Wu~\cite{Yan_06}
indicate that at RH 100\%, the net water flux through
the membrane is close to $\alpha_w j_0$, where $\alpha_w \simeq 0.2$.
To improve stability of the numerical algorithm, the static water flux was fixed
at $\alpha_w j_0$, meaning that the second boundary condition to Eq.\eqref{eq:lam0x} was
\begin{equation}
   \left(- A_{\lam} \tD^0 \pdr{\lam^0}{\tx}\right)_{\tx=-\tl_m}  + n_d\tj^0 = \alpha_w \tj_0
   \label{eq:lam0x_lbc}
\end{equation}

Integrating Eq.\eqref{eq:lam0x} from $- \tl_m$ to $\tx$, we get the first-order equation
expressing conservation of the diffusive water flux in the membrane
\begin{equation}
    \tD^0 \pdr{\lam^0}{\tx} = p, \quad \lam^0|_{\tx=0-} = \Lambda(RH^c) + \alpha_s s^0|_{\tx=0+}
    \label{eq:lam0x1}
\end{equation}
where
\begin{equation}
    p = \left.\tD^0 \pdr{\lam^0}{\tx}\right|_{\tx=-\tl_m}.
    \label{eq:pdef}
\end{equation}

The parameter $p$ can be determined from Eq.\eqref{eq:lam0x_lbc}:
\begin{equation}
   p = \dfrac{(n_d  - \alpha_w)\tj_0}{A_{\lam}}
   \label{eq:p0}
\end{equation}
With Eq.\eqref{eq:p0}, equation \eqref{eq:lam0x1} forms a Cauchy problem.
If necessary, it can be solved backward, from $\tx=0$ to $\tx = -\tl_m$,
using a standard Runge-Kutta method. However, the impedance problem requires
only the membrane water content and its first derivative at $\tx=0$ (see below).
These values can be calculated directly from the equations of this section:
\begin{multline}
   \lam^0|_{\tx=0-} = \Lambda(RH^c) + \alpha_s s^0|_{\tx=0+}, \\
   \left.\pdr{\lam^0}{\tx}\right|_{\tx=0-} = \dfrac{p}{\tD^0(\lam^0|_{\tx=0-})}
   \label{eq:lamx0}
\end{multline}


\section{Model for the cathode side impedance}

The core of the cathode side impedance model with the two-phase
water transport in the CCL
was developed by Sun \etal\cite{Kulikovsky_24b}.
The model equations are listed in Table~\ref{tab:eqs}; for the
notations see the Nomenclature section.
The cell performance model \cite{Kulikovsky_24b}
includes through-plane equations for the oxygen, proton
and water transport, Eqs.\eqref{eq:cbx}--\eqref{eq:sx}.
\begin{table}
\small
\begin{tabular}{|p{5.8cm}|p{2.3cm}|}
\hline
\multicolumn{2}{|c|}{Transport equations} \\
\hline
\begin{equation}
   \pdr{c_h}{t} + v\pdr{c_h}{z} = - \dfrac{D_b}{h}\left.\pdr{c_b}{x}\right|_{x=l_t+l_b}
   \label{eq:chz}
\end{equation} & $$\text{Channel}$$ \\
\hline
\begin{equation}
   \pdr{c_b}{t} - D_b\pddr{c_b}{x} = 0
   \label{eq:cbx}
\end{equation} & $$\text{GDL}$$ \\
\hline
{\begin{align}
   &\pdr{c}{t} - \pdr{}{x}\left(D_{ox}\pdr{c}{x}\right) = -\dfrac{Q_{ORR}}{4F} \label{eq:cx} \\
   &C_{dl}\pdr{\eta}{t} - \pdr{}{x}\left(\sion\pdr{\eta}{x}\right) = - Q_{ORR} \label{eq:etax} \\
   &\dfrac{1}{V_l}\pdr{s}{t} - \dfrac{K_l}{V_l \mu_l}\pddr{p_l}{x}
       = \dfrac{Q_{ORR}}{2F}   \label{eq:sx}
\end{align}}   &  $$\text{CCL}$$ \\
\hline
\multicolumn{2}{|c|}{Source function} \\
\hline
\begin{equation}
   Q_{ORR} = i_*\left(\dfrac{c}{\cref}\right)\lexp{\dfrac{\eta}{b}}
\label{eq:QORR}
\end{equation} &  $$\text{ORR rate}$$ \\
\hline
\multicolumn{2}{|c|}{Water retention curve (WRC) and} \\
\multicolumn{2}{|c|}{the CCL transport coefficients} \\
\hline
\begin{multline}
   s = s_0  \\
   + \dfrac{(1 - s_0)}{2}
             \left(1 + \tanh\left(\dfrac{p_c - p_{c,0}}{k_{pc}}\right) \right)
\label{eq:sat}
\end{multline} & $$\text{WRC}$$ \\
\begin{equation}
   \sion = \ksig s
\label{eq:sion}
\end{equation} & $$\text{H$^+$ conductivity}$$ \\
\begin{equation}
   \Dox = D_{ox,d}(1 - s)^3
   \label{eq:Dox}
\end{equation} & $$\text{O$_2$ diffusivity}$$ \\
\hline
\end{tabular}
\caption{Eqs.\eqref{eq:chz}--\eqref{eq:cx}, the governing equations
    for oxygen, water and proton transport. GDL stands for the gas diffusion layer.
    Eq.\eqref{eq:QORR} is     the ORR rate.
    Eqs.\eqref{eq:sat},\eqref{eq:sion}, \eqref{eq:Dox} -- the WRC, proton conductivity vs saturation,
    and the oxygen diffusivity vs saturation, respectively.
    For the notations see Nomenclature section.
}
\label{tab:eqs}
\end{table}

A model extension that takes
into account the convective air transport in the channel, Eq.\eqref{eq:chz},
is reported in Reshetenko \etal\cite{Kulikovsky_24e}.
This extension is particularly important at the low air flow stoichiometry
of $\lam_c=2$ used in this work, as the local current and oxygen distributions
along the channel are strongly nonuniform.
The local current density distribution along the channel
was calculated using the model\cite{Kulikovsky_19a}.
It is worth noting that in~\cite{Kulikovsky_19a,Kulikovsky_24e} the quadratic
dependence of the CCL effective oxygen diffusivity $\Dox$ on the parameter $(1-s)$ in
Eq.\eqref{eq:Dox} was used.
Here, however, we used the cubic dependence, as with the scaling $(1-s)^2$ we were unable
to achieve a satisfactory fit to the low-frequency part of the spectra below 1 Hz.

The boundary conditions express the continuity of oxygen concentration
and flux at the channel/gas diffusion layer (GDL) and GDL/CCL interfaces, as well as
the influx of liquid water from the membrane to the CCL (see below).
They also represent zero oxygen flux in the membrane and zero proton
current at the CCL/GDL interface. Further details can be
found in~\cite{Kulikovsky_24b,Kulikovsky_24e}.
The 1d+1d performance model in Table~\ref{tab:eqs}
was linearized and Fourier-transformed to obtain
the linear boundary-value problem for the small perturbation amplitudes of the
oxygen concentration and ORR overpotential. In this work,
we set the water vapor concentration in the CCL equal to the saturated value, $c_v = c_v^{sat}$.
This is equivalent to zero evaporation/condensation rates.

The $5\times5$ cm$^2$ meander flow field was modeled as a single straight channel
divided into eight segments. The through-plane equations were linked to the
oxygen transport equation in the channel using a 1d+1d approach.
In each segment, solving the through-plane problem yields
the local segment impedance $Z_{seg}$:
\begin{equation}
   Z_{seg} = - \left. \dfrac{\eta^1}{\sion\pdra{\eta^1}{x}}\right|_{x=0}
   \label{eq:Zseg}
\end{equation}
where
$x=0$ corresponds to the membrane interface,
$\eta^1$ is the overpotential perturbation and
$\sion$ is the CCL proton conductivity.
The segments are connected in parallel and the total cell impedance was calculated as
\begin{equation}
   Z_{cell} = \left(\dfrac{1}{8}\sum_{n=1}^8\dfrac{1}{Z_{seg, n}} \right)^{-1}
              + \ri\omega L_{cab} S_{cell} + R_{HFR}
   \label{eq:Zcell}
\end{equation}
where $L_{cab}$ is the cable inductance, $S_{cell}$ is the cell active area, and $R_{HFR}$ is the
high-frequency (ohmic) cell resistance.

\section{Models coupling and numerical aspects}

The membrane  water content $\lam$ and water flux must be coupled
with the saturation and water flux in the CCL.
Consider first the steady-state solutions in the membrane
and CCL. The dimensionless static equation for the liquid pressure $\tp_L^0$ in the CCL,
Eq.\eqref{eq:sx}, reads
\begin{multline}
   - \tk_L \pddr{\tp_L^0}{\tx} = 2 \tp_{O_2}^0 \expo, \\
   - \left.B_L\pdr{\tp_L^0}{\tx}\right|_{\tx=0+} = \alpha_w \tj_0, \quad
       \tp_L^0|_{\tx=1} = \tp_{cell}
   \label{eq:tpLx}
\end{multline}
where the boundary condition at $\tx = 0$ fixes the static water flux
entering from the membrane to the CCL. Here,
\begin{equation}
   \tk_L = \dfrac{4 F p_{O_2} K_L}{i_* \lcat^2 \mu_w V_w}
   \label{eq:tkL}
\end{equation}
and the other parameters appearing in Eq.\eqref{eq:BLA} are described in the Nomenclature section.

The problem for the small perturbation amplitude $\tp_L^1$ at the membrane/CCL
interface \cite{Kulikovsky_24b} reads
\begin{multline}
   \tk_L \pddr{\tp_L^1}{\tx} = \gamma^2\ri\tom s^1
       - 2\expo\left(\tpox^1 + \tpox^0\teta^1\right), \\
   \shoveleft{- \left.B_L\pdr{\tp_L^1}{\tx}\right|_{\tx=0+}} \\
          = \left(- A_{\lam} \left(\tD^0 \pdr{\lam^1}{\tx}
              + \pdr{D^0}{\lam}\pdr{\lam^0}{\tx}\lam^1 \right) + n_d\tj^1\right)_{\tx=0-}, \\
           \tp_L^1|_{\tx=1} = 0
   \label{eq:tp1x}
\end{multline}
where the first boundary condition means continuity of the water flux perturbations
on either side of the membrane/CCL interface.
Here, $\gamma$ is the constant parameter
\begin{equation}
   \gamma = \sqrt{\dfrac{4 F}{V_w\Cdl b}}, \quad
   \label{eq:gamma}
\end{equation}

Using Eq.\eqref{eq:dbcFlam1} in the first boundary condition to Eq.\eqref{eq:tp1x}, this
condition is reformulated in terms of the perturbation amplitude
of CCL liquid saturation $s^1$:
\begin{multline}
  - \left.B_L\pdr{\tp_L^1}{\tx}\right|_{\tx=0+} \\
             = \left(- A_{\lam}\alpha_s \left(\tD^0 \pdr{s^1}{\tx}
                 + \pdr{D^0}{\lam}\pdr{\lam^0}{\tx} s^1 \right) + n_d\tj^1\right)_{\tx=0-},
   \label{eq:dtpLdx_bc}
\end{multline}
Eq.\eqref{eq:dtpLdx_bc} closes the problem for the small perturbation
amplitudes in the CCL. The relaxation of the membrane water content appears
in the model through the coefficients in Eq.\eqref{eq:dtpLdx_bc},
which are derived from Eq.\eqref{eq:lamx0}.
Eq.\eqref{eq:dtpLdx_bc} is the only correction to the model for the
perturbation amplitudes discussed in Sun \etal\cite{Kulikovsky_24b}.
A complete formulation of the latter problem can be found in Appendix D
of Ref.\cite{Kulikovsky_24b}.

The custom fitting code was developed using the Python SciPy library
and the MPI library for parallel calculations.
Solution of the static problems for the membrane, Eq.\eqref{eq:lam0x1},
and CCL, Eqs.\eqref{eq:tpLx}, does not require iterations. The
static membrane problem
was solved using the Runge-Kutta solver {\em solve\_ivp}.
The linear boundary-value problems (BVPs) for
the perturbation amplitudes and the nonlinear BVPs for the static
variables (pressures and overpotential) were solved using the {\em solve\_bvp}
SciPy solver.

Eq.\eqref{eq:Zcell} was fitted to the experimental spectra using the
SciPy least-squares module {\em least\_squares}.
Fitting of a single spectrum with 96 frequency points takes around five minutes on 48 cores
of a parallel cluster with Intel Xeon processors,
or around half an hour on an eight-core notebook.

\section{Experimental}

The experimental work was performed at Hawaii Natural Energy Institute (HNEI) using
a custom-built fuel cell test station and a proprietary impedance spectroscopy system.
Catalyst coated membranes with a geometric area of 25 cm$^2$ were supplied by Gore.
The Pt loading was 0.4 mg$_{Pt}$ cm$^{-2}$ for both the anode and cathode. The membrane electrode
assembly (MEA) utilized 25 BC GDLs for both electrodes and Teflon gaskets
with a thickness of 150 $\mu$m, providing a 20\% compression ratio.

A 25 cm$^2$ single cell hardware was manufactured at Fuel Cell Technology Inc. The flow field design
featured a triple serpentine configuration for both the anode and cathode, with channels
0.787 mm wide and 1.02 mm deep, and ribs 0.787 mm wide.
A co-flow gas arrangement was employed throughout this study.

The anode and cathode feed gases were humidified to the desired level, and the cell was operated
at a backpressure of 1.5 bar and a temperature of 80$^\circ$C. Minimum gas flow rates were fixed
to provide a stoichiometry of 2 at 200 mA cm$^{-2}$. Thus, the reactant stoichiometry was 2 for H$_2$
and air at current densities above 200 mA cm$^{-2}$, while at 100 mA cm$^{-2}$ the stoichiometry
was increased to 4. The operating conditions are summarized
at Table~\ref{tab:oper}.

EIS was conducted under galvanostatic load control in a frequency range of 10 kHz to 0.1 Hz
using 20 points per decade. The  amplitude of current perturbation was selected such that
the amplitude  of cell voltage oscillations remained below 10 mV.

\begin{table}
\small
\begin{tabular}{|l|c|}
\hline
   	 Cathode pressure, bar               & $1.5$ \\
     Anode/cathode   RH, \%   & 32/32, 50/50, 100/100 \\
     Cell temperature, K                 & $273 + 80$ \\
     Hydrogen/air flow stoichiometry     & 2/2  \\
                                         & (2/4 for 100 mA~cm$^{-2}$) \\
\hline
     CCL thickness $\mu$m                & 10   \\
     GDL thickness $\mu$m                & 235 \\
\hline
\end{tabular}
\caption{The regime of cell operation and the estimated thicknesses of the CCL and GDL.}
\label{tab:oper}
\end{table}

\section{Results and discussion}

\begin{figure}
\begin{center}
   \includegraphics[scale=0.45]{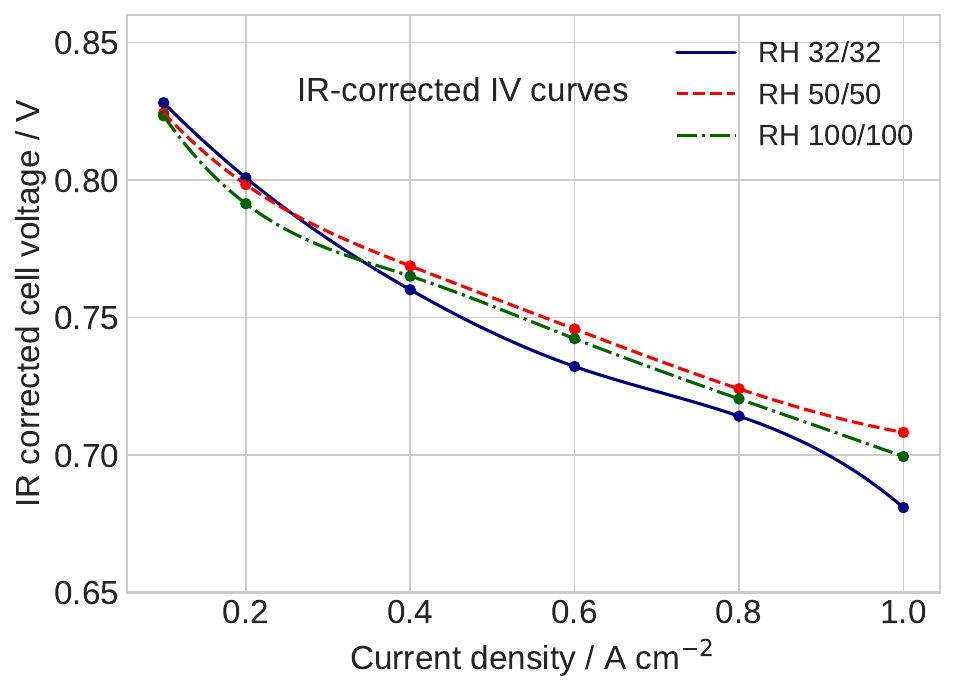}
\caption{The experimental (points) and interpolated (lines) IR-corrected polarization curves
      of the cell operated at the indicated relative humidities. The points are
      interpolated using cubic spline just as an eye guide.
  }
\label{fig:ivc}
\end{center}
\end{figure}

Figure~\ref{fig:ivc} shows the IR-corrected polarization curves of the cell
operating with the anode/cathode relative humidity of 32/32\%, 50/50\% and 100/100\%.
These curves were measured at the current densities corresponding to
those at which the impedance spectra were acquired. As can be seen,
the difference between the IR-corrected cell potentials does not exceed
15 mV (Figure~\ref{fig:ivc}). One might expect the impedance spectra
measured at the same currents and different RH values to produce similar
cell parameter values, except for high-frequency cell resistivity.

\begin{figure}
\begin{center}
   \includegraphics[scale=0.55]{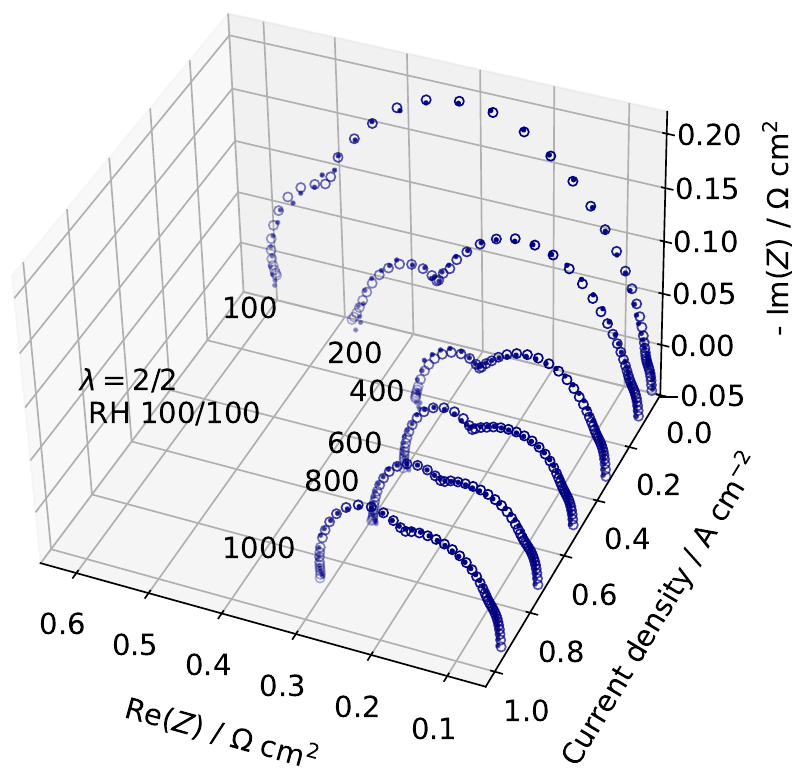}
\caption{The experimental spectra (points) and fitted model Eq.\eqref{eq:Zcell} (open circles)
   for the indicated current densities (mA~cm$^{-2}$).
   To improve readability, half of the points (10 per decade) are shown.
   The regime of cell operation is indicated in Table~\ref{tab:oper}, RH 32/32\%.
  }
\label{fig:3d}
\end{center}
\end{figure}

\begin{figure}
\begin{center}
   \includegraphics[scale=0.5]{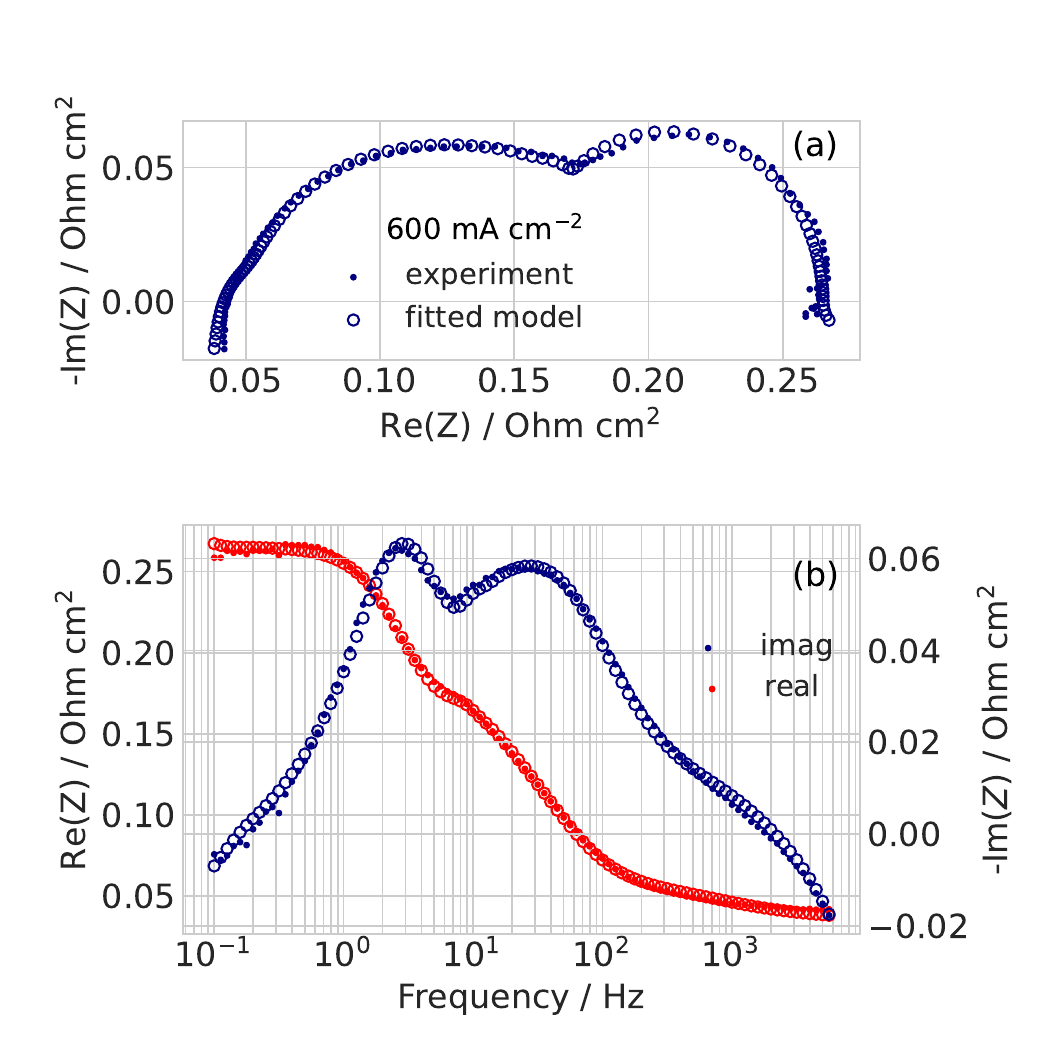}
\caption{The experimental spectra (points) and fitted model Eq.\eqref{eq:Zcell} (open circles)
   for the current density of 600 mA~cm$^{-2}$ and RH 100/100\%.
  }
\label{fig:600}
\end{center}
\end{figure}

Figure~\ref{fig:3d} shows the experimental and fitted model
Nyquist spectra for RH 100/100\%. The model fits the spectra well.
The spectra for RH 32/32\% and 50/50\% are fitted
equally well. A more detailed plot of the experimental and fitted model
spectra is shown in Figure~\ref{fig:600}.


\begin{figure*}
\begin{center}
    \includegraphics[scale=0.45]{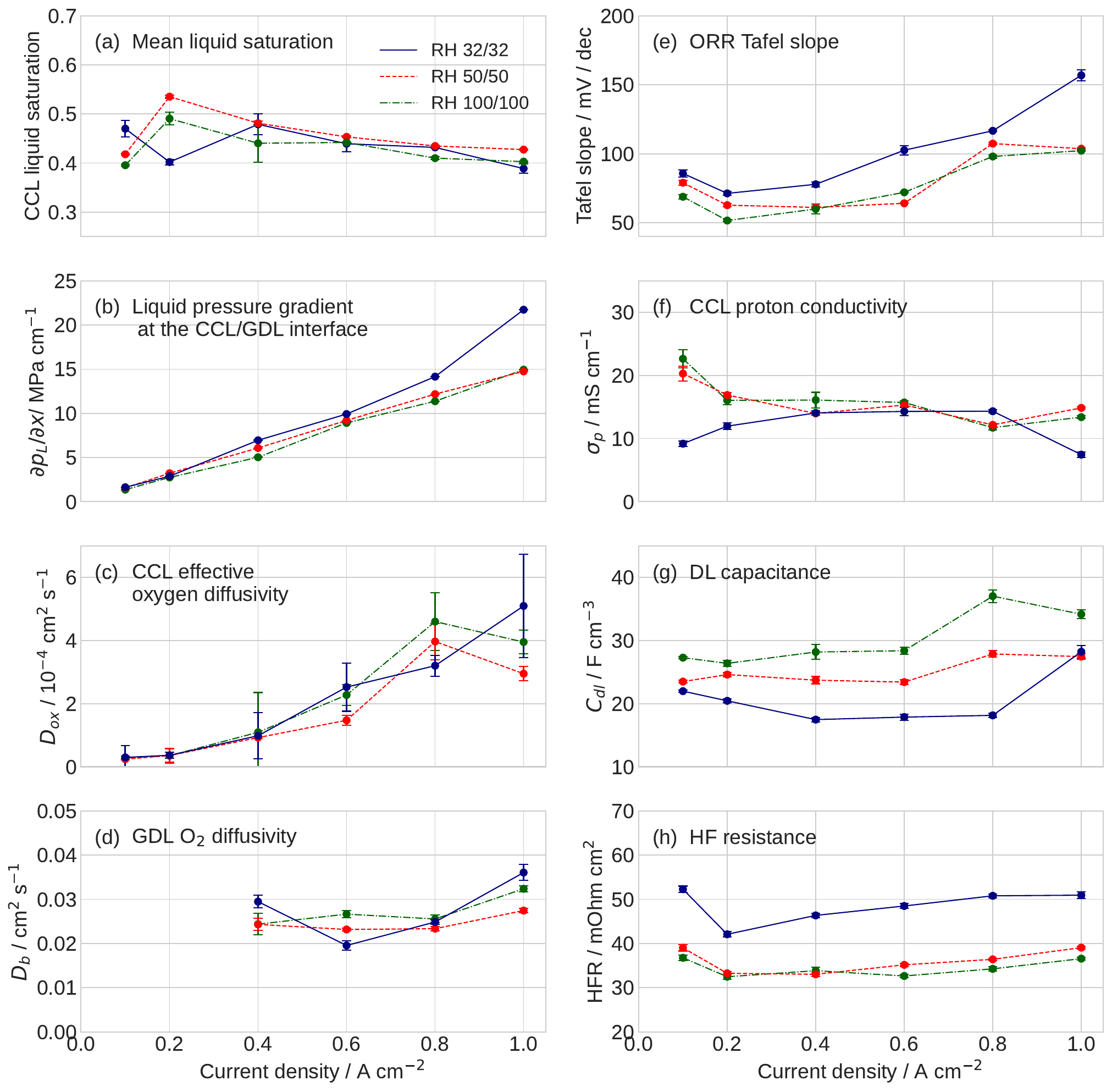}
\caption{The dependence of fitted parameters on the cell current density
    at RH 32/32\%, 50/50\%, and 100/100\%. The lines are only intended as a guide.
    }
\label{fig:parms}
\end{center}
\end{figure*}

Figure~\ref{fig:parms} shows the dependence of the fitted cell parameters
on the current density. The error bars in Figure~\ref{fig:parms} were calculated
from the Jacobian matrix returned by the {\em least\_squares} procedure.
The size of the error bar indicates how sensitive the fitted spectrum
is to variation in the respective parameter.
A  small bar indicates high sensitivity and vice versa.

In general, the curves for different RH demonstrate similar
trends, with important variations discussed below. This similarity gives us
confidence that the model correctly captures key phenomena in the cell.

Contrary to common belief, the CCL mean liquid saturation decreases with the cell
current density for all three RH values (Figure~\ref{fig:parms}a).
This decrease is due to the liquid pressure gradient in the CCL,
which increases linearly with the cell current and removes liquid
water from the catalyst layer (Figure~\ref{fig:parms}b).

\begin{figure}
\begin{center}
    \includegraphics[scale=0.45]{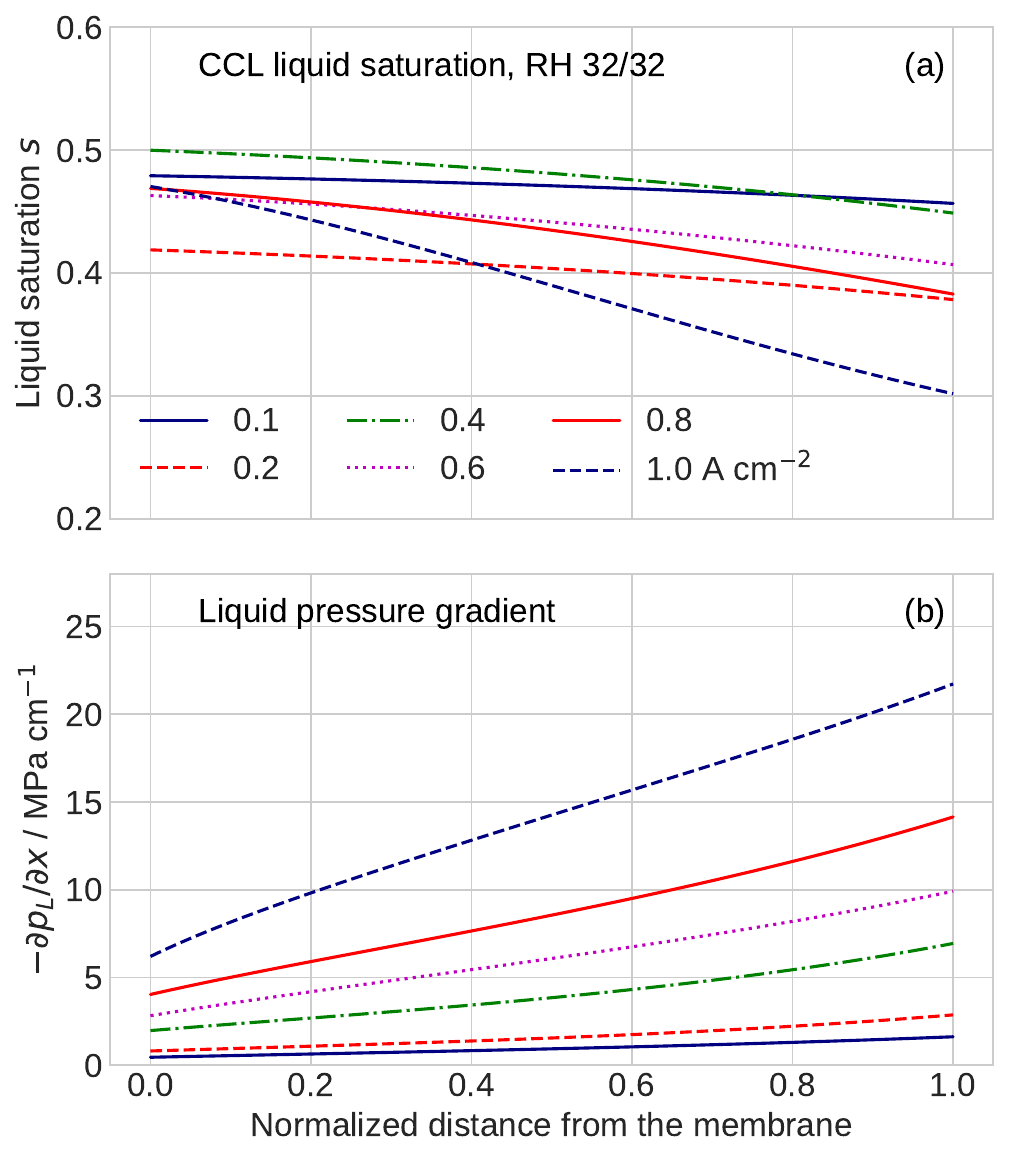}
\caption{The shapes of (a) the liquid saturation and (b) the (negative) liquid pressure
      gradient through the CCL depth for  RH 32/32\% and the indicated current densities
      (A~cm$^{-2}$).
    }
\label{fig:sx32}
\end{center}
\end{figure}

\begin{figure}
\begin{center}
    \includegraphics[scale=0.45]{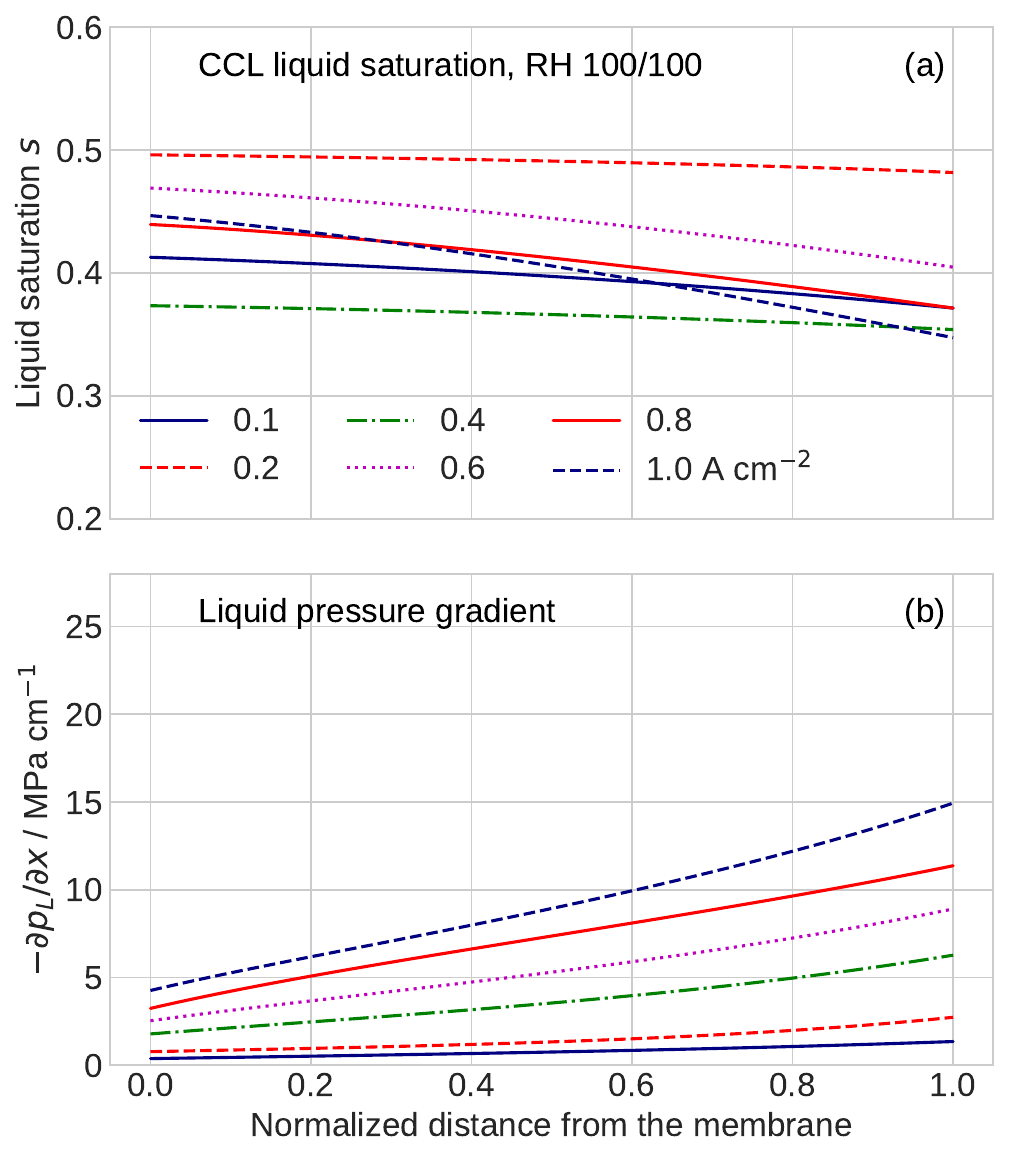}
\caption{The shapes of (a) the liquid saturation and (b) the (negative) liquid pressure
      gradient through the CCL depth for  RH 100\%/100\% and the indicated current densities
      (A~cm$^{-2}$).
    }
\label{fig:sx100}
\end{center}
\end{figure}

The through-plane shapes of the liquid saturation and the liquid pressure
gradient for RH 32/32\% and RH 100/100\% are shown in Figures~\ref{fig:sx32} and
\ref{fig:sx100}, respectively. The effect of reducing $s$
at currents 0.8~A~cm$^{-2}$ (solid red curve) and 1.0~A~cm$^{-2}$
(dashed blue curve) is clearly visible. Furthermore, at RH 32/32\% and
high currents, the saturation is much less uniform along the $\tx$--coordinate
(Figure~\ref{fig:sx32}a).


At all RH values,  the CCL effective oxygen diffusivity $\Dox$ exhibits
rapid growth with the cell current density  (Figure~\ref{fig:parms}c).
This growth is partly due to the decreasing liquid saturation, see Eq.\eqref{eq:Dox}.
However, the most significant factor contributing to this growth
is the progressive involvement of larger pores in the process of current
conversion. As discussed by Reshetenko \etal\cite{Kulikovsky_24e}, a small current is converted
in the smallest pores due to their
large surface area. The Knudsen oxygen diffusivity of small pores is proportional to their radius.
As the current grows, larger pores become involved in the current conversion
process, leading to an increase in the effective diffusivity as measured by EIS.

The in-situ parameter $\Dox$ measured here should not be confused with the
ex-situ CCL diffusivity measured, for example, by Suzuki \etal\cite{Suzuki_20b} using
a microfluidic device. In the ex-situ experiment \cite{Suzuki_20b}, the measured diffusivity
of gaseous oxygen through the void CCL pores is
more than two orders of magnitude higher than the CCL oxygen diffusivity shown
in Figure~\ref{fig:parms}c. In an operating PEM fuel cell, the oxygen transport pathway
to Pt surface includes the ionomer film barrier covering the Pt/C agglomerates
and the diffusion inside the agglomerates, which are filled with water.
These transport elements dramatically reduce the effective oxygen diffusivity
in the operating CCL as compared to the ex-situ values.

The GDL oxygen diffusivity $D_b$ varies between 0.02 and 0.04 cm$^2$~s$^{-1}$
depending on the RH value (Figure~\ref{fig:parms}d). At currents
of 100 and 200 mA~cm$^{-2}$, the contribution of oxygen transport
in the GDL to the cell impedance is negligible and cannot be captured by the model.

The ORR Tafel slope increases  with the cell current density from 60 to 150 mV/decade
(Figure~\ref{fig:parms}e). At the lowest RH 32/32\%, the Tafel slope is notably
higher than at the other two RH combinations, suggesting that liquid water
accelerates some of the reaction steps in the ORR.

The CCL proton conductivity $\sion$ varies from 10 to 20 mS~cm$^{-1}$,
with notably lower values at RH 32/32\% (Figure~\ref{fig:parms}f).
At RH 50/50\% and 100/100\%, $\sion$ decreases with the cell current, following
the CCL liquid saturation.
Note the significant decay of $\sion$ at RH 32/32\% and a cell current density
of 1000 mA~cm$^{-2}$ (Figure~\ref{fig:parms}f, the blue curve). This decay,
together with the increasing ORR Tafel slope (Figure~\ref{fig:parms}d)
is responsible for the bending of the blue polarization curve in Figure~\ref{fig:ivc}.

The parameters sensitive to relative humidity are the DL capacitance, which increases
with RH (Figure~\ref{fig:parms}g) and the high-frequency (ohmic)
cell resistivity, which decreases with RH (Figure~\ref{fig:parms}h).
The positive effect of water on DL capacitance suggests that water increases
the electrochemically active surface area (ECSA).
This behavior correlates with the results of CO stripping experiments
of Shinozaki, Yamada and Morimoto \cite{Shinozaki_11},
which indicate a steady increase in Pt utilization as RH changes from 25\% to 100\%.

The high-frequency (ohmic) cell resistance depends weakly on the cell current
and strongly on the RH (Figure~\ref{fig:parms}h). The key factor here is the
{\em anodic} RH. The large HFR resistance at RH 32/32\% (Figure~\ref{fig:parms}h)
shows that the liquid water back diffusion
from the CCL to the anode side is small, meaning that the membrane can only be
humidified from the anode side. Indeed, the effective diffusivity of liquid water in
the Nafion membrane decreases as the membrane water content falls (Figure~\ref{fig:Dm}).
Thus, the electroosmotic drag of water drying out the anode side of the membrane
simultaneously lowers the water diffusivity there. Once the anode side of the membrane
is dry, it cannot be humidified by water back diffusion from the cathode
side\cite{Kulikovsky_03a}. This effect is the main reason why, in practical applications,
the anode flow RH is typically kept above 50\%.

\begin{figure}
\begin{center}
    \includegraphics[scale=0.45]{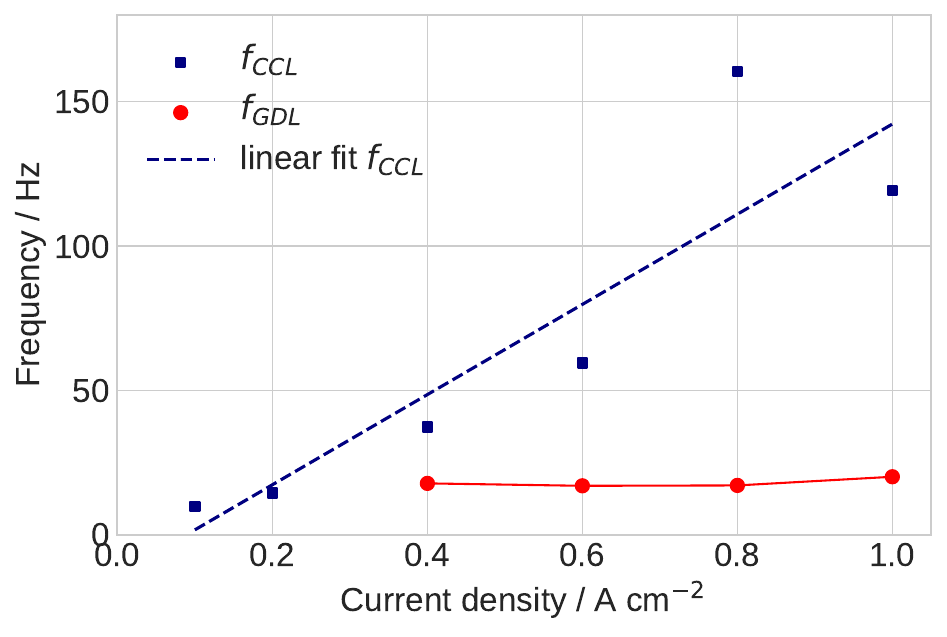}
\caption{The characteristic frequencies of the oxygen transport
    in the CCL (blue squares and the dotted line) and in the GDL
    (red filled circles and the solid line) vs the cell current density for RH 100/100\%.
    }
\label{fig:freq}
\end{center}
\end{figure}

Finally, the data in Figures~\ref{fig:parms}c,d allow us
to compare the characteristic frequencies of the oxygen transport in the GDL $f_{GDL}$
and the CCL $f_{CCL}$. For the estimation we use the Warburg finite-length formula for the
characteristic frequency $f$ of mass transport in the porous layer
\begin{equation}
   f = \dfrac{0.404 D}{l^2},
   \label{eq:fmt}
\end{equation}
where $D$ is the oxygen diffusivity of the porous layer and $l$ is the layer thickness.
The result is shown in Figure~\ref{fig:freq}. As can be seen, $f_{CCL}$ is close to $f_{GDL}$
at the current density of 100 mA~cm$^{-2}$. At this current density, the DRT would produce
a single oxygen transport peak representing the O$_2$ transport in both the layers.
However, as the current density increases,  $f_{CCL}$ increases
linearly, while $f_{GDL}$ remains almost constant at around 10 Hz.
At a cell current of 1~A~cm$^{-2}$,  $f_{CCL}$ is an order of magnitude larger than  $f_{GDL}$
(Figure~\ref{fig:freq}). Thus, as the cell current increases, the frequency $f_{CCL}$ overlaps with
the ORR charge-transfer frequency, and may even exceed it.
This effect should be taken into account in the DRT analyses of PEMFC spectra.

\section{Conclusions}

We developed a model of the membrane impedance in a PEM fuel cell.
This model was then coupled
with a recent two-phase model for the cell cathode impedance.
We fitted the complete model to measured spectra of a PEMFC operated
at three relative humidities of of 32/32\%, 50/50\% and 100/100\%.
Six spectra were acquired for each RH combination
at current densities ranging from 100 to 1000 mA~cm$^{-2}$. The dependence of the fitting
parameters on the cell current shows the following trends.
\begin{itemize}

\item The mean CCL liquid saturation $\bars$
    decreases with the cell current in the range of 600 to 1000 mA~cm$^{-2}$.

\item The decrease in liquid saturation is due to the growing liquid pressure gradient,
   which removes water from the CCL.

\item The CCL effective oxygen diffusivity $\Dox$ increases dramatically with the cell current.
   Following Ref.\cite{Kulikovsky_24e}, we attribute this growth to the progressive
   involvement of larger pores to the proton current conversion.

\item The ORR Tafel slope increases with the cell current from 60 to 150 mV/decade.
   At RH 32/32\%, the Tafel slope is significantly higher than at the other two RH combinations.

\item The CCL proton conductivity $\sion$ is about 15 mS~cm$^{-1}$. At RH 50/50\% and 100/100\%,
    it decreases with the cell current. At RH 32/32\%, $\sion$ shows
    no clear correlation with the current density.

\item The double layer capacitance increases with the RH indicating growth
    of the ECSA with the RH. This effect correlates with the experiments~\cite{Shinozaki_11}.

\item The GDL oxygen diffusivity, which is in the range of
   0.02 to 0.04 cm$^2$~ s$^{-1}$ is almost independent of the current density.

\item At RH 32/32\%, the high-frequency cell resistivity is significantly higher, than at
    RH 50/50\% and 100/100\%. This indicates poor wetting of the anode side of the membrane
    by water back diffusion from the cathode side. We attribute this effect to
    the reduction in ionomer water diffusivity as the water content decreases.

\end{itemize}

\section{Acknowledgments}

T. Reshetenko gratefully acknowledges funding from the US Office
   of Naval Research (N00014-25-1-2372).

\appendix

\section{Equation for the perturbation amplitude of the membrane water content}

To linearize Eq.\eqref{eq:lamx} we substitute into Eq.\eqref{eq:lamx} $D_w = D_w^0(x) + D_w^1(x,t)$,
$\lam = \lam^0(x) + \lam^1(x,t)$, neglect products of the small
terms and subtract the static equation for $\lam^0$.
For the small-amplitude transient $\lam^1(x,t)$ we find the equation
\begin{multline}
   \pdr{\lam^1}{t} - \left(2 \pdr{D^0}{\lam} \pdr{\lam^0}{x}\right)\pdr{\lam^1}{x} \\
       -  \left(\pdr{\lam^0}{x}\right)^2 \pdr{D^1}{\lam}
       - D^1\pddr{\lam^0}{x} - D^0 \pddr{\lam^1}{x} = 0
   \label{eq:lam1x}
\end{multline}
Using $D^1 = \left(\pdra{D^0}{\lam}\right) \lam^1$,
$\pdra{D^1}{\lam} = \left(\pddra{D^0}{\lam}\right) \lam^1$, and substituting
into Eq.\eqref{eq:lam1x} the Fourier-transform
\begin{equation}
   \lam^1(x,t) = \lam^1(x,\omega) \exp(\ri\omega t),
\end{equation}
we get the equation for the perturbation amplitude $\lam^1(x,\omega)$:
\begin{multline}
   D^0 \pddr{\lam^1}{x}
     + \left(2 \pdr{D^0}{\lam} \pdr{\lam^0}{x} \right)\pdr{\lam^1}{x} \\
          +  \left(\left(\pdr{\lam^0}{x}\right)^2 \pddr{D^0}{\lam} + \pdr{D^0}{\lam}\pddr{\lam^0}{x} - \ri\omega \right) \lam^1
           = 0
   \label{eq:Flam1x2}
\end{multline}
It is convenient to transform Eq.\eqref{eq:Flam1x2}
using the dimensionless variables \cite{Kulikovsky_24b}.
Multiplying Eq.\eqref{eq:Flam1x2} by $4 F \cref \lcat^2 / (\sigma_* b)$, we get
Eq.\eqref{eq:dFlam1x2}.



\section*{Nomenclature}

\small

\begin{tabular}{ll}
    $\tilde{}$   &  Marks dimensionless variables                          \\
    $\Alam$      &  Dimensionless parameter, Eq.\eqref{eq:BLA}              \\
    $a_w$        &  Water vapor activity                                   \\
    $B_L$        &  Dimensionless parameter, Eq.\eqref{eq:BLA}              \\
    $b$          &  ORR Tafel slope, V                                     \\
    $c$          &  Oxygen molar concentration in the CCL, mol~cm$^{-3}$   \\
    $c_b$        &  Oxygen molar concentration in the GDL, mol~cm$^{-3}$   \\
    $c_h$        &  Oxygen molar concentration in the channel, mol~cm$^{-3}$   \\
    $\cref$      &  Reference oxygen molar concentration, mol~cm$^{-3}$   \\
    $c_w$        &  Water molar concentration in the membrane,  mol~cm$^{-3}$        \\
    $c_v^s$      &  Saturated water concentration at 80°C                  \\
    $\Cdl$       &  Double layer volumetric capacitance, $F$~cm$^{-3}$      \\
    $D_b$        &  Oxygen diffusivity of the GDL, cm$^2$~s$^{-1}$         \\
    $D_w$        &  Water diffusivity of the membrane, cm$^2$~s$^{-1}$         \\
    $\Dox$       &  Oxygen diffusivity of the CCL, cm$^2$~s$^{-1}$         \\
    $D_{ox,d}$   &  Oxygen diffusivity of the dry CCL, cm$^2$~s$^{-1}$     \\
    ${\rm e}_0$  &  Elementary charge, $C$                                 \\
    $F$          &  Faraday constant, C~mol$^{-1}$                         \\
    $h$          &  Cathode channel height, cm                             \\
    $\ri$        &  Imaginary unit                                         \\
    $i_*$        &  ORR volumetric exchange current density, A~cm$^{-3}$   \\
    $j_0, j^0$   &  Static current density, A~cm$^{-2}$                    \\
    $K_L$        &  CCL liquid water permeability, cm$^{2}$       \\
    $k_{pc}$     &  Slope of the water retention curve, Eq.\eqref{eq:sat}, Pa       \\
    $k_{\sigma}$ &  Coefficient in Eq.\eqref{eq:sion} S~cm$^{-1}$       \\
    $k_L$        &  Liquid water transport coefficient, mol cm$^{-1}$ Pa $^{-1}$ s$^{-1}$    \\
    $L_{cab}$    &  Cable inductance, H \\
    $l_b$        &  GDL thickness, cm                                         \\
    $l_m$        &  Membrane thickness, cm                                     \\
    $\lcat$      &  CCL thickness, cm                                         \\
    $N_w$        &  Water flux in the membrane, mol~cm$^{-2}$~s$^{-1}$        \\
    $n_d$        &  Water drag coefficient in the membrane                   \\
    $p$          &  Dimensionless parameter, Eq.\eqref{eq:pdef}   \\
    $p_c$        &  Capillary pressure, Pa               \\
    $p_g$        &  Gas phase pressure, Pa               \\
    $p_L$        &  Liquid phase pressure, Pa                \\
    $p_{O_2}$    &  Oxygen partial pressure, Pa                \\
    $p_{cell}$   &  Cathode pressure, Pa                      \\
    $Q_{ORR}$    &  ORR rate, A~cm$^{-3}$                              \\
    $R$          &  Gas constant, J~K$^{-1}$~mol$^{-1}$                \\
    $R_{HFR}$    &  High frequency cell resistivity, Ohm cm$^2$  \\
    $s$          &  CCL liquid saturation     \\
    $s_0$        &  Minimal CCL liquid saturation, Eq.\eqref{eq:sat}     \\
    $t_*$        &  Characteristic time of DL charging, s \\
    $V_w$        &  Liquid water molar volume, cm$^3$~mol$^{-1}$   \\
    $W_m$        &  Ionomer equivalent weight, g~mol$^{-1}$    \\
    $v$          &  Air flow velocity in the channel, cm~s$^{-1}$            \\
    $x$          &  Coordinate through the cell, cm       \\
    $z$          &  Coordinate along the channel, cm       \\
    $Z_{seg}$    &  Segment impedance, Ohm~cm$^2$       \\
    $Z_{cell}$   &  Cell impedance, Ohm~cm$^2$       \\[1em]
\end{tabular}

{\bf Subscripts:\\}

\begin{tabular}{ll}
    $0$      & membrane/CCL interface \\
    $b$      & GDL  \\
    $ref$    & Reference value \\
    $h$      & Channel  \\[1em]
\end{tabular}

{\bf Superscripts:\\}

\begin{tabular}{ll}
    $0$      & Steady--state value \\
    $1$      & Small--amplitude perturbation \\ [1em]
\end{tabular}

{\bf Greek:\\}

\begin{tabular}{ll}
    $\alpha_s$          &  Dimensionless parameter in Eq.\eqref{eq:lamx} \\
    $\alpha_w$          &  Net water transfer coefficient in membrane \\
    $\eta$              &  ORR overpotential, positive by convention, V \\
    $\Lambda$           &  Membrane water sorption isotherm   \\
    $\lam$              &  Membrane water content \\
    $\lam_c$            &  Air flow stoichiometry \\
    $\mu_w$             &  Liquid water kinematic viscosity, Pa~s \\
    $\rho_m$            &  Mass density of a dry ionomer, g~cm$^{-3}$ \\
    $\sion$             &  CCL proton conductivity, S~cm$^{-1}$        \\
    $\sigma_*$          &  Reference proton conductivity, S~cm$^{-1}$        \\
    $\omega$            &  Angular frequency of the AC signal, s$^{-1}$
\end{tabular}

\newpage

\end{document}